\documentclass[12pt]{iopart}

\usepackage{graphicx} 
\usepackage{bbm}
\usepackage{amssymb}

\begin{document}

\title[Social Aggregation as a Cooperative Game]{Social Aggregation as a Cooperative Game}

\author{Daniele Vilone$^1$ and Andrea Guazzini$^{2}$}
\address{$^1$ Center for the Study of Complex Dynamics, Universit\`a di Firenze, Florence, Italy;
Departamento de Matem\'aticas, Universidad Carlos III de Madrid, Legan\'es (Madrid), Spain}
\address{$^2$ Institute for Informatics and Telematics, National Research Council (CNR), Pisa, Italy}
\ead{daniele.vilone@gmail.com}

\begin{abstract}
A new approach for the description of phenomena of social aggregation is suggested.
On the basis of psychological concepts (as for instance social norms and cultural coordinates),
we deduce a general mechanism for the social aggregation in which different clusters of
individuals can merge according to the cooperation among the agents. In their turn, the agents
can cooperate or defect according to the clusters distribution inside the system. The fitness
of an individual increases with the size of its cluster, but decreases with the work the
individual had to do in order to join it.
In order to test the reliability of such new approach, we introduce a couple of simple
toy models with the features illustrated above.
We see, from this preliminary study, how the cooperation is the most convenient strategy
only in presence of very large clusters, while on the other hand it is not necessary to have
one hundred percent of cooperators for reaching a totally ordered configuration with only one
megacluster filling the whole system.
\end{abstract}


\section{Introduction}

The study of the evolution of social systems is a topic nowadays attracting the interest
of researchers from
different domains such as physics, psychology, mathematics. In fact, an interdisciplinary approach
provides a more powerful way to understand and model such complex systems~\cite{revcast}.
One important issue within this field is the understanding of the phenomena of social aggregation,
as for instance urbanization, cultural clusterization, imitative processes in econophysics.

The classical approach of sociophysics is by means of Statistical Mechanics: the system
under analysis is considered in a thermodynamical way, {\it i.e.\rm} it is
seen as composed by a great number of identical elementary units and, starting
from the rules governing the microscopical dynamics of individuals, the general
behaviour at macroscopical level is achieved. Consequently, this methodology is
very useful in those systems whose peculiarity is produced by statistical laws
rather than by specific microscopic details~\cite{deff00,bagnoli07,cliff73,holl75,vilo04,axel97}.

On the other hand, a different approach is also possible, by means
of the analysis of cooperative behaviours, in particular the study
of the emergence of cooperation in systems of generic agents~\cite{axel81,nowak92,jime08},
in financial markets~\cite{li08} or in academic networks~\cite{pugliese08}. The main
theoretical scaffolding to face such issues is borrowed from game theory, largely
used in econophysics, which focuses on the evolution of the strategies that agents
use during their interactions~\cite{neum,friedm}.


Social norms, beliefs, attitudes and opinions are also concepts which have attracted
the interest of researchers from a great number of different
fields~\cite{mengele,chalub06,kolstad07,tuu08}. Certainly, it is
quite hard to define explicitly those objects, but reaching a reliable representation
of them is a required step in order to implement models for social dynamics.
Psychology and sociology are useful tools to provide definitions for concepts like
previous ones, but what we need here is just an operational definition in order to make
our models useful.

On the one hand, sharing the same beliefs, attitudes, opinions and in general social norms,
means to use the same ``cultural coordinates'' to communicate, enhancing the
process of ``social meaning negotiation''~\cite{foster98}, defined as the interaction of
two individuals who do not share the same lexicon or meanings, and increasing the
probability to converge to the same "social cluster".
On the other hand, social fragmentation can be viewed also as the result of this same
process~\cite{foster98}.
Moreover, social norms are objects intrinsically linked together
and there is a natural resistance to change cultural coordinates because it is possible to
see them as  the product of well-established neural circuits and because frequently a
change would cause a cascade effect on the others.
Thus, importing a psychological representation of cultural coordinates (CC) means at
least to take into consideration three main characteristics:

\begin{itemize}
\item CC are hard to change.

\item People who have the same CC belong to the same cultural cluster.

\item The degree of cultural separation among agents, that is how much their social
norms are different, can be defined as a sort of "distance" in the abstract space
of the CC.
\end{itemize}

The last crucial ingredient we have to consider here is the role of the
environment on the negotiation strategies. Indeed, from a sociological
point of view, it is well-known that belonging to a big cultural cluster
({\it i.e.\rm} sharing the same CC with
a great amount of people) increases the individual fitness~\cite{gira93,cald02}.
Consequently the macroscopic features of a population influence the probability
to change its own CC to increase the size of the group.

Finally, the main role of social sciences in this challenge is to link in an ecological
way the microscopic dynamics ({\it i.e.\rm} the evolution strategy of individual) with the
macroscopic phenomenology ({\it i.e.\rm} the state of the whole system).

\section{Social aggregation and game theory}
\label{model}

The purpose of this paper is to study the phenomena of social aggregation by unifying the
thermodynamical approach with the game-theoretical one, and using the psychological concepts
depicted in the Introduction.
More precisely, we want to write down models whose microscopical dynamics is defined
starting from the payoff matrix of each agent. In other words, the interaction between
two individuals is determined (also) by their payoff matrix, and in their turn the
payoff matrices of the individuals evolve according with the dynamics. The details of
the dynamics, {\it i.e.\rm} the payoff matrix, will be determined on the basis of
psycho-sociological considerations.
We stress that our aim is just to suggest a new methodology, therefore the models
introduced here have the minimum amount of refinement required for such a purpose.

Let us consider a system of $N$ agents where every agent belongs to a cluster. Each cluster
represents a group of individuals who share the same cultural coordinates. From a
psychological point of view we start considering two main assumptions. First, we assume
that an agent tends to maintain his CC. At the same time every agent has an advantage to
belong to a group as big as possible. On the basis of such considerations we can state
that the fitness of an individual increases with the number of other individuals sharing
its same social norms, {\it i.e.\rm} with the size of its cluster. On the other hand, the fitness
decreases according to an ``economic criterion'', that is according to the work the
individual accomplished in order to merge with its actual group. In practice, when a player
$i$ meets an opponent $j$ from a stranger cluster, its payoff matrix is

\begin{equation}
\label{paymatr0}
\hat A= \left\{A_{\kappa\lambda}\right\}=\left(
\begin{array}{cc}
 w_1(m_j)-\frac{w_2(d_{ij})}{2}\ \ \  & \ \ \ w_1(m_j-m_i+1)-w_2(d_{ij}) \\ w_1(1) & 0
\end{array}
\right)
\end{equation}

where the indices $\kappa,\ \lambda$ can indicate $C$, ``cooperation'' (availability to
join the opponent's group), $D$, ``defection'' (that is ``no cooperation''), while $m_i$
is the population of the cluster of the player $i$, $m_j$ the population of the
opponent's cluster, and $d_{ij}$ is the distance (in the CC space) between the
two clusters. Finally, the function $w_1(m)$ is the fitness contribution of $m$ individuals,
and $w_2(d)$ is the work ({\it i.e.\rm} the loss of fitness) an agent has to bear to cover
a distance equal to $d$.

The meaning of Eq. (\ref{paymatr0}) is then the following: when two cooperators meet,
they put in common
their CC, and this is equivalent to the merging of their groups into one. Moreover,
we assume that they ``meet in the middle'', so that the work spent is half of the one
due to the original distance between them. If the opponent does not cooperate, the first
player has to cover the entire distance $d_{ij}$ to gain the CC of the opponent's
cluster, and he will lose contact with its original group. On the contrary, if the first
player does not cooperate (but the opponent does) it will not spend anything but gain to
its group only the presence of the second player. Finally, if nobody cooperates, nothing
will happen. Of course, it is meant that only players from different clusters can meet,
or equivalently, that when two players from the same group meet, nothing happens.

It must be noticed that the property of the clusters to merge when two cooperators meet
is a strong assumption.
Anyway, there are many situations in which this assumption is quite realistic. For example,
let us consider the spreading of technological or cultural advances through different
populations: when an individual meets from a stranger group a new technique useful to face
successfully some not yet resolved problem, presumably he will import that into his original
social cluster. If the new technique improves appreciably the fitness of the population, then
it will soon become a common knowledge of all the members, and under this aspect the two
clusters have merged together. Similar processes can happen for other cultural instances,
as for example languages, religions, traditions.
Finally, we stress the fact that the
goal of this paper is just to present a couple of simple toy models in order to show how
this new approach should work. Thus, the models we are going to present in the following
sections, both developed starting from Eq. (\ref{paymatr0}), are very simple and deserve
to be improved in the future.

At last, for simplicity it is reasonable to assume the fitness functions $w_1(m)$ and $w_2(d)$
as proportional to the population and to the distance, respectively. Moreover, to make
easier calculations, we set equal to 1 the proportionality constants ($\Rightarrow w_i(x)=x$),
so that the payoff matrix gets the general form

\begin{equation}
\label{paymatr1}
\hat A=\left(
\begin{array}{cc}
 m_j-\frac{d_{ij}}{2}\ \ \  & \ \ \ m_j-m_i+1-d_{ij} \\ 1 & 0
\end{array}
\right)
\end{equation}

\section{Static Homogeneous Model}
\label{hom}

As a first step our study, we analyze an oversimplified static model, which
we call ``static homogeneous model'' (SHM): we assume that the system is always
perfectly homogeneous, so that all the players obey to the same payoff matrix.
More precisely, we consider a system made up by a great number of identical clusters,
each one of the same size $m$ and at the same distance from each other:
$d_{ij}=2x(1-\delta_{ij})$.
Moreover, we consider such distance $2x$ big enough to consider the fitness contribution of
one individual negligible with respect to the work needed to cover it: $2x\gg1$ (anyway, as
it is easy to verify, this approximation does not change appreciably the physics of the model).
So, the payoff matrix $\hat A$ of Eq. (\ref{paymatr1}) becomes

\begin{equation}
\label{homA}
\hat A_H= \left(
\begin{array}{cc}
1+\varepsilon \ \ \  & \ \ \ -2x \\ 1 & 0
\end{array}
\right)
\end{equation}

where $\varepsilon=m-x-1$ (this will turn out to be the crucial parameter of the SHM).
Of course it is always $x>0$. Now, we want just to understand which is the rational strategy
the agents should adopt when meeting foreigners (interactions between players of the same
clusters are not taken into account), in the given configuration, neglecting any possible
time evolution.

Said $p(t)$ the density of cooperators, its behaviour is given by
replicator equation~\cite{sigmund,nowak}

\begin{equation}
\frac{\dot p}{p}=f_C-\langle \hat A_H\rangle
\end{equation}

where $f_C$ is the averaged payoff of a cooperator and $\langle \hat A_H\rangle$ the
averaged payoff of a generic player. We have to stress the fact that we consider this
system as frozen, and the time evolution given by the previous equation must be seen
just as a mathematical trick in order to discover the Nash equilibria of the matrix
(\ref{homA}). Explicitly, the replicator equation becomes

\begin{equation}
\label{rplc}
\frac{dp}{dt}=(2x+\varepsilon)\cdot p(1-p)(p-\omega)
\end{equation}

with

\begin{equation}
\label{omga}
\omega=\frac{2x}{2x+\varepsilon}
\end{equation}

The Nash equilibria of Eq. (\ref{rplc}) are in general the roots of the polynomial
at right side:

\begin{equation}
\label{equilibri}
\left\{
\begin{array}{rl}
p_1^E=0 & \ \ \ \ \ \ \ \ \ \ \ \ \ \ \ \ \mbox{(no cooperators)} \\
p_2^E=1 & \ \ \ \ \ \ \ \ \ \ \ \ \ \ \ \ \mbox{(all cooperators)} \\
p_3^E=\omega & \ \ \ \ \ \ \ \ \ \ \ \ \ \ \ \ \mbox{(mixed equilibrium)}
\end{array}
\right.
\end{equation}

In order to understand the phenomenology, it is important to find also the stability
of the equilibria given in Eq. (\ref{equilibri}). The explicit evaluation of the
stability is left in Appendix, here we just give the results obtained.

CASE $\varepsilon<0$ - The payoff matrix is here a Prisoner Dilemma's one.
This condition is equivalent to $m<x+1$: the distance among clusters
is high enough that the work needed to merge with another group is always
greater than the maximum gain possible in case of cooperation.
Thus, in this case the Nash equilibrium $p_1^E=0$ is the only one which is stable:
$p_2^E=1$ is unstable and $p_3^E$ is not physical, since $\omega>1$.

\begin{figure}[h]
\begin{center}
\includegraphics[angle=0,width=8cm,clip]{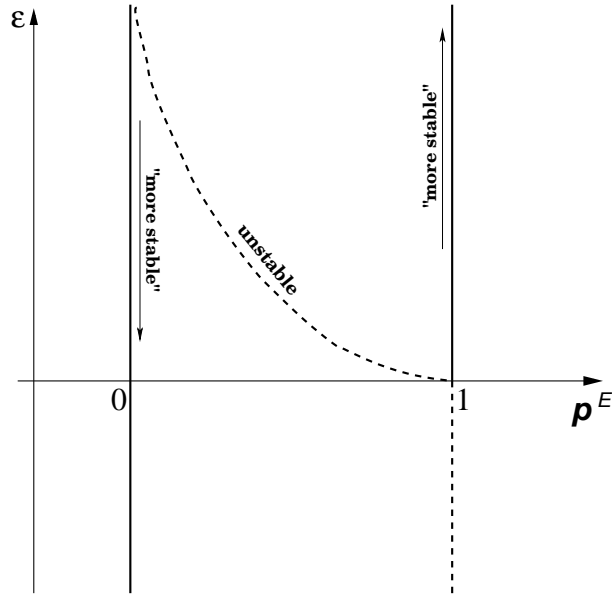}
\end{center}
\caption{
Phase diagram of the equilibria of the system described by Eq. (\ref{rplc}). For $\varepsilon>0$
the continuous lines at $p^E=0$
and $p^E=1$ represent the (pure) stable Nash equilibria, the dotted line represents the (mixed) unstable
Nash equilibrium $p^E=\omega=2x/(2x+\varepsilon)$. For $m\rightarrow (x+1)^+$,
{\it i.e.\rm} $\varepsilon\rightarrow0^+$ the unstable equilibrium collapses on $p^E=1$,
while for $m\rightarrow+\infty$, {\it i.e.\rm} $\varepsilon\rightarrow+\infty$
it collapses on $p^E=0$: in this limit only the all-cooperators configuration is stable.
For $\varepsilon<0$ $p_1^E$ remains stable, whilst $p_2^E$ is unstable: there is actually a
bifurcation in $(p^E,\varepsilon)=(1,0)$.}
\label{phasediagr}
\end{figure}

CASE $\varepsilon>0$ - Now we have a Stag Hunt payoff matrix. In this case all the
three equilibria given in Eq. (\ref{equilibri}) are physical. More precisely,
pure equilibria $p_1^E$ and $p_2^E$ are stable, while the mixed equilibrium $p_3^E$ is unstable. Because now it is
$m>x+1$, the gain in fitness in case of mutual cooperation is bigger than the loss due to the distance, so that
the stable equilibrium $p_2^E$ is perfectly understandable. The fact that the
equilibrium $p_1^E$ is stable also in this case could be a little bit surprising, but a deeper analysis
of the situation gives back a more intuitive picture: the basin of attraction of equilibrium
$p_2^E$ increases for $m$ (and then $\varepsilon$) increasing, while at the same
time the basin of $p_1^E$ decreases, disappearing in the limit $\varepsilon\rightarrow+\infty$.
In this sense, we could state that for great values of $\varepsilon$ the equilibrium $p_2^E$
is somehow ``more stable'' than $p_1^E$, and viceversa for small values of $\varepsilon$.
More precisely, the basin of attraction of $p_2^E$ becomes bigger than the basin of $p_1^E$
(that is, the all-cooperators equilibrium becomes ``more stable'' than the no-cooperators
one), when it is $m>3x+1$.
A phase diagram of the SHC is given in Figure \ref{phasediagr}.

Despite its roughness, this simple model allows us to draw some preliminary conclusions.
In particular, it seems to be clear that cooperation is an advantageous
strategy only when the size of the clusters is much bigger than their averaged distance.
In the next section we will improve our investigation by means of the dynamical
homogeneous model.

\section{Dynamical Homogeneous Model}
\label{simul}

The main feature of the SHM is that the system is frozen, {\it i.e.\rm} does not evolve in
time: we set it in a given configuration (a great number of equal clusters of the same size
and at the same distance from each other) and wonder which is the most rational strategy
agents should adopt in order to improve their own fitness, without making them ``play the
game'' for real.
What we want to do now is to write down a model with the general properties stated in
section \ref{model}, which can also evolve dynamically in time. For this purpose, we
are now going to introduce the ``dynamical homogeneous model'' (DHM).

DHM is implemented as follows. At $t=0$ we divide a system of $N$ individuals into clusters
each one of size $m_0$, so that we have initially $N/m_0$ clusters of the same size
(we always set $N$ as a multiple of $m_0$).
Every generic cluster $i$ is identified by a natural variable
$g_i\in \{1,2,...,\frac{N}{m_0}\}$: then, the distance between two agents belonging to
the clusters $j$ and $k$ respectively will be $d_{jk}=|g_j-g_k|$ (notice that with such
definition the distance is a discrete variable too). Moreover, each agent
has a default strategy (cooperative or not cooperative),
picked up randomly, so that the initial density of cooperators is $\varrho_0$.
The dynamics works in this way: at each elementary step two different agents, $i$ and $j$,
are drawn. If they belong
to the same cluster, nothing happens. Otherwise, they ``play the game'' according
to the payoff matrix (\ref{paymatr1}) and their actual strategy: if
both players cooperate, their clusters merge (the smallest is absorbed by the biggest
one); if one player defects, the cooperator leaves its cluster and joins the opponent's
one; if nobody cooperates nothing happens. After the game, a player computes what
it would have gained if it had adopted the other strategy (remaining fixed the
strategy of the opponent). If such virtual payoff is greater than the real one,
the player will change its strategy at the next interaction.
For simplicity, in order to have easier simulations, even though
the payoff is always calculated by means of the matrix (\ref{paymatr1}), in case
of two clusters merging (when a pair of cooperators from different groups meet), the smallest
group enters the biggest one: in practice, they spend fitness as "meeting in the middle",
but in fact, it is the small cluster to reach the big one in its position.
Time is measured in montecarlo steps, so that on average every agent interacts once
per time unit. We accomplished all our simulations with $\varrho_0=1/2$ and for several
values of $m_0$ and $N$.

\begin{figure}[h]
\begin{center}
\includegraphics[angle=0,width=8cm,clip]{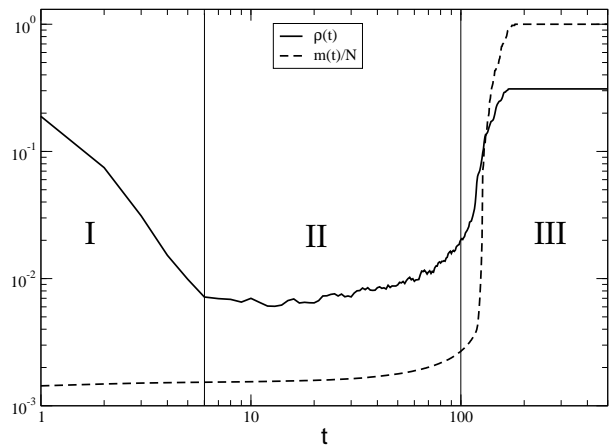}
\end{center}
\caption{
Plot of the behaviour of the cooperators density (tick line) and of
the averaged size of the survived
clusters divided by $N$ (dashed line) for a DHM system with $N=3024$, $m_0=4$ and
$t_{max}=500$ time units. Data averaged after 25 simulations. The three different
dynamical regimes are clearly distinguishable.}
\label{gendyn}
\end{figure}

In Figure \ref{gendyn} we report the typical behaviour of the DHM for a particular
choice of the parameters ($N=3024, m_0=4$). This figure well summarizes the phenomenology
of our model. We can clearly distinguish three different dynamical regimes: at early
times we have Regime I, that we also call ``exponential decay regime'' for reasons we will
soon explain, then we find a steady-state regime or  Regime II, and finally we have
Regime III, in which the system rapidly reaches a frozen state: we are going to study them
separately in the following subsections.

Before analysing in details the three dynamical regimes, it is convenient to write
down the equations ruling the evolution of the main quantities which characterize the
state of the system. Concerning the cooperators density, which will be here indicated by
$\varrho(t)$, starting from the payoff matrix
$\hat A$ written in Eq. \ref{paymatr1}, it is easy to see that its time evolution must
be ruled by the equation

\begin{equation}
\label{rhogen}
\frac{d\varrho}{dt}=\left\langle \frac{N-m_i}{N}\left[-2\beta_{ij}\varrho^2+(1-\beta_{ij}-
\alpha_{ij})\varrho(1-\varrho)+2(1-\alpha_{ij})(1-\varrho)^2\right]\right\rangle_{i,j}
\end{equation}

where $i$ is an agent randomly extracted, $j$ another agent randomly extracted not belonging
to the same cluster of $i$, $m_i$ the size of the cluster of $i$, $\beta_{ij}$ the probability
that $m_j-d_{ij}/2$ is smaller than one, $\alpha_{ij}$ the probability that the quantity
$m_j-m_i+1-d_{ij}$ is smaller than zero. Finally, the symbol $\langle\cdot\rangle_{i,j}$
means of course the average over every possible couple $i,j$ (with $i$ and $j$ belonging
to different clusters). Analogously, the time evolution of the averaged size of the
survived clusters, $m(t)$, will be given by

\begin{equation}
\label{mgen}
\frac{dm}{dt}=\left\langle\frac{m_i(N-m_i)}{N^2}\cdot m_j\varrho^2\right\rangle_{i,j}
\end{equation}

\subsection{Exponential decay regime}
\label{RI}

At the very early stages of the dynamics, we can assume that the payoff matrix of
each agent (when interacting with foreigners) has the form

\begin{equation}
\label{dhm0}
\hat A_0=\left(
\begin{array}{cc}
 m_0-\frac{d_{ij}}{2}\ \ \  & \ \ \ 1-d_{ij} \\ 1 & 0
\end{array}
\right)
\end{equation}

with $m_0\simeq m_i$ $\forall i$ and, as we have already stated, $d_{ij}=|g_i-g_j|$.
In such case we have $\beta_{ij}=\beta$ $\forall i$ and $\alpha_{ij}=1$ $\forall i ,j$:
in the limit $N\gg m_i$ (we will treat the case of $m_0$ equal to a finite fraction of
N in subsection \ref{GVm}) equations (\ref{rhogen}) and (\ref{mgen}) become

\begin{equation}
\label{rhoI}
\dot{\varrho}(t)=-\beta\varrho(\varrho+1)
\end{equation}

and

\begin{equation}
\label{mI}
\dot{m}(t)=\frac{m^2\varrho^2}{N}
\end{equation}

whose solutions are, respectively

\begin{equation}
\label{solrhoI}
\varrho(t)=\frac{\varrho_0e^{-\beta t}}{1+\varrho_0(1-e^{-\beta t})}\simeq\varrho_0e^{-\beta t}
\end{equation}

and

\begin{equation}
\label{solmI}
m(t)=\frac{4N\beta m_0}{4N\beta-m_0(1-e^{-2\beta t})}
\end{equation}

\begin{figure}[h]
\begin{center}
\includegraphics[angle=0,width=8cm,clip]{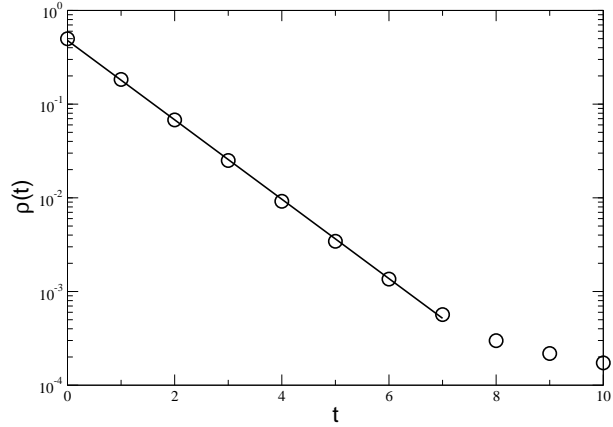}
\end{center}
\caption{
Plot of the initial behaviour of the cooperators density as a function of time for $m_0=1$,
$N=10000$ after 100 simulations. The empty circles are the numerical data, the full line is
the exponential fit $\sim\exp(-\beta t)$, with $\beta=0.99\pm0.01$.}
\label{expD1}
\end{figure}

\begin{figure}[h]
\begin{center}
\includegraphics[angle=0,width=8cm,clip]{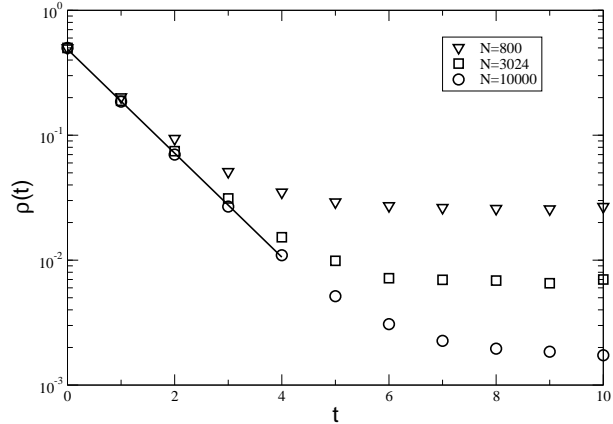}
\end{center}
\caption{
Plot of the initial behaviour of the cooperators density as a function of time for $m_0=4$,
$N=800$ (triangles), $N=3024$ (squares) and $N=10000$ (circles) after 100 simulations.
The full line is the exponential fit $\sim\exp(-\beta t)$ for $N=10000$, with
$\beta=0.97\pm0.01$.}
\label{expD2}
\end{figure}

Now, said $d=\langle d_{ij}\rangle_{i,j}$, in this regime it is
$\beta=\Pr(m_0-d/2<1)$, and this probability depends in general
on $m_0$ and $N$. Anyway, it is straightforward to understand that $\beta(m_0=1)=1\ \forall N$,
and that $\lim_{N\rightarrow+\infty}\beta(m_0)=1\ \forall m_0$. On the basis of these
considerations, we expect an exponential decay of $\varrho(t)$ at early stages of
dynamics, with coefficient $\beta$ equal to one for $m_0=1$, and tending to one for
increasing values of the size $N$ of the entire system if $m_0>1$. This fact is fully confirmed
by Figures \ref{expD1} and \ref{expD2}.


\begin{figure}[h]
\begin{center}
\includegraphics[angle=0,width=8cm,clip]{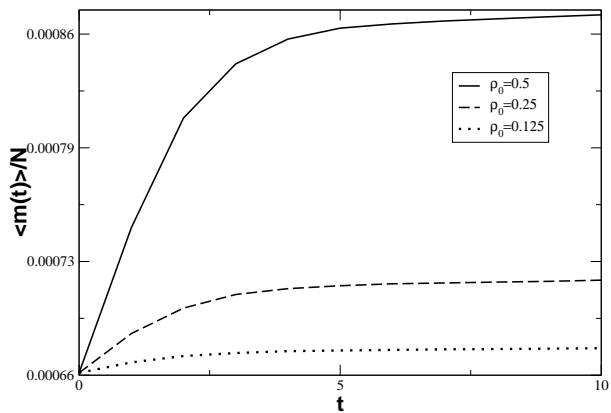}
\end{center}
\caption{
Initial increasing (up to $t_{max}=10$) of averaged clusters size (divided by $N$), for
a system with $N=3024,\ m_0=2$ and $\rho_0=0.5$ (full line), $\rho_0=0.25$ (dashed line),
$\rho_0=0.125$ (dotted line); data averaged over 25 different simulations. The small
increasing step ($m(t_{max})/m_0\sim1.3$ for $\rho_0=0.5$) takes place at the very early
stages (until $t_{max}\simeq3$), and rapidly decreases with $\rho(0)$ decreasing.}
\label{inincr}
\end{figure}

Regarding the averaged size of survived clusters, we see from Eq. (\ref{solmI})
that, while it remains valid, $m(t)$ is bigger than $m_0$ and smaller than the quantity
\[
m^{\infty}=\left( \frac{1}{m_0}-\frac{1}{4N\beta}\right)^{-1}
\simeq m_0\left(1+\frac{m_0}{4N\beta}\right)\simeq m_0
\]

so that $m(t)$ is practically constant during this regime: a proof
of the last statement is given already in Figure \ref{gendyn}, where
it is clear how $m(t)$ is a quasi-constant in the initial stages of
the dynamics. More precisely, it is a quasi-constant apart a small
initial increasing due to the interactions among cooperators during
the very early times of the dynamics, and indeed such increase
vanishes in the limit $\rho_0\rightarrow0^+$ (see Figure
\ref{inincr}). It must be noticed that, because for every survived
cluster $k$ it must be $m_k>0$, and we are dealing with small values
of $m_0$, this means that the clusters distribution inside the
system remains rather close to the initial one (see also Figure
\ref{cldstr}). Moreover, we emphasize the fact that the quantity
$m^{\infty}$ is just a limit superior
 of $m(t)$ during the exponential decay regime, and not a value that the averaged size
can effectively reach.

\

The exponential decay regime will last until the cooperators density is not too small:
we expect actually that it should end when $\varrho$ becomes of the order of $N^{-1}$.
From Eq. (\ref{solrhoI}) we gain

\begin{equation}
\label{tstar1}
\varrho_0e^{-\beta t^*}\approx\frac{1}{N}\ \Longrightarrow\
t^*\approx\frac{1}{\beta}\log(\varrho_0N)
\end{equation}

For the case depicted in Figure \ref{expD1} ($\beta=1,\ \varrho_0=0.5,\ N=10000$), previous
relation gives $t^*\approx8.5$, in good agreement with the numerical data. For values
of $m_0$ greater than 1, the evaluation of $t^*$ directly from Eq. (\ref{tstar1}) is more
delicate because in this case also the quantity $\beta$ depends in its turn on $N$, and
moreover there are bigger fluctuations in the system (when two cooperators meet their
groups merge, and this causes bigger fluctuations in clusters distribution as $m$
increases); however the relation $t^*\propto\log(N)$ is valid $\forall m_0$, as we will
see in subsection \ref{RII}. Therefore, for $N\rightarrow+\infty$ this regime never ends:
$\varrho(t)\rightarrow0$ and, from Eq. (\ref{solmI}), we find $m(t)\simeq m_0=\mbox{const}$.
Then, in the thermodynamical limit (when $m_0\ll N$ for every finite $m_0$) we have a
similar result of the SHM, where the unique (stable) Nash equilibrium is the complete
absence of cooperators. On the other hand, this is coherent with the fact that, if we
set the system {\it ab initio\rm} with $\varrho_0=0$, nothing will ever happen.

\subsection{Steady state}
\label{RII}

Once, for finite values of $N$, the cooperators density became very small and
the system left the Regime I, the equation (\ref{rhoI}) is not valid any more.
Indeed, in this case almost every interaction will be between two defectors,
so that equation (\ref{rhogen}) becomes

\begin{equation}
\label{rhoII}
\dot\varrho(t)\simeq-2(1-\alpha)(1-\varrho)^2
\end{equation}

where we took into accounts that from Eq. (\ref{solmI}) the clusters distribution
is practically the initial one, and then we assumed again $N\gg m_i$ and
$\alpha_{ij}=\alpha\ \forall i,j$. But, as we have just said, the clusters density
is still almost equal to the initial one, so that it must be also $\alpha=\Pr(1-d<0)\simeq1$,
from which

\begin{equation}
\label{solrhoII}
\dot\varrho(t)\simeq0\ \Longrightarrow\ \varrho(t)\simeq\varrho_{ss}=\mbox{const.}
\end{equation}

\begin{figure}[h]
\begin{center}
\includegraphics[angle=0,width=8cm,clip]{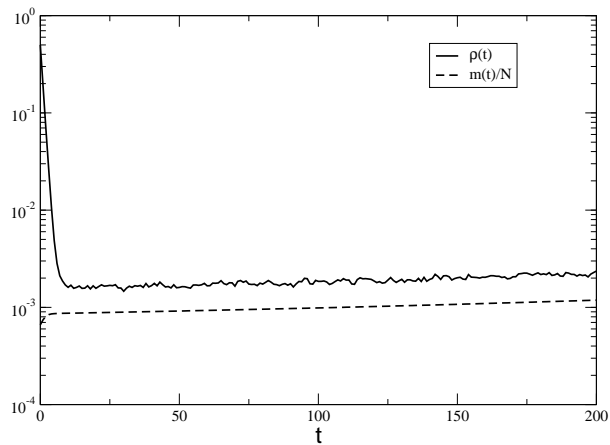}
\end{center}
\caption{
Plot of the cooperators density and of the normalized size of survived clusters for
$N=3024$ and $m_0=2$ after 100 simulations. After the initial decay, there is a clear
steady state regime in which both $\varrho(t)$ and $m(t)/N$ remain almost constant.}
\label{stst1}
\end{figure}

\begin{figure}[h]
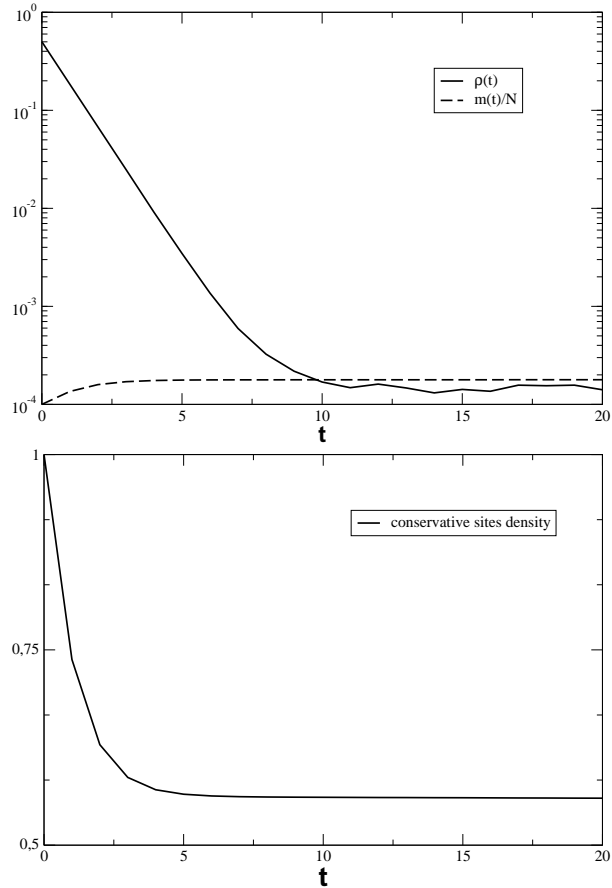

\begin{center}
\includegraphics[clip=true,width=8cm]{rhoL10000m1+.eps}
\includegraphics[clip=true,width=8cm]{konsL10000m1+.eps}
\end{center}
\caption{Upper graph: cooperators density and averaged clusters size (divided by $N$) for
a DHM system with $N=10000$ and $m_0=1$ until $t_{max}=20$ time units (data averaged after 100
different simulations). Lower graph: density of ``conservative sites'', that is the sites
which are still in the initial cluster, for the same system of the upper graph. As one can
see, after an initial drop, the majority of agents (about the $60\%$) is still in its original
group also during the steady state regime.}
\label{cldstr}
\end{figure}

In Figure \ref{stst1} we can see this behaviour for the case $N=3024$ and
$m_0=2$; in Figure \ref{cldstr} we show instead how the majority of the agents
remains in the initial cluster also during this second dynamical regime.

As we can easily see, $\varrho(t)$ is actually almost constant, just slightly
increasing because of small fluctuations in the clusters distribution which
make $\alpha$ not perfectly equal to one, but a very little bit smaller.
On the other hand, $m(t)$ keeps on behaving as in the exponential decay regime. This can be
seen by inserting Eq. (\ref{solrhoII}) into (\ref{mI}), obtaining

\begin{equation}
\label{solmII}
\dot m(t)=\frac{\varrho_{ss}^2}{N}m^2(t)\ \Longrightarrow\
m(t)\simeq\frac{m_0N}{N-m_0\varrho_{ss}^2t}
\end{equation}

where we exploited again the fact that $m(t)$ does not change too much during the Regime I.
Now, while the quantity $m_0\varrho_{ss}^2t$ remains much smaller than $N$, also $m(t)$
remains close to $m_0$ (see Figures \ref{gendyn} and \ref{stst1}). However,
once the relation $m_0\varrho_{ss}^2t\ll N$ ceases to be true, the system exits from
the steady state regime, because at this point $m(t)\gg m_0$ and the clusters
distribution is now quite different from the initial one: so, also the quantities
$\alpha_{ij}$ in Eq. (\ref{rhogen}) become considerably smaller than 1 and this
changes dramatically the shape of $\varrho(t)$ too, as shown in Figure \ref{gendyn}.

Before starting the analysis of the subsequent regime, it is worth to take a look
to the behaviour of $\varrho_{ss}$ as a function of $N$ and $m_0$. For $m_0=1$, from
Eq. (\ref{tstar1}) it has to be necessarily

\[
\varrho_0e^{-\beta t^*}\equiv\varrho_{ss}\sim\frac{1}{N}
\]

\begin{figure}[h]
\begin{center}
\includegraphics[angle=0,width=8cm,clip]{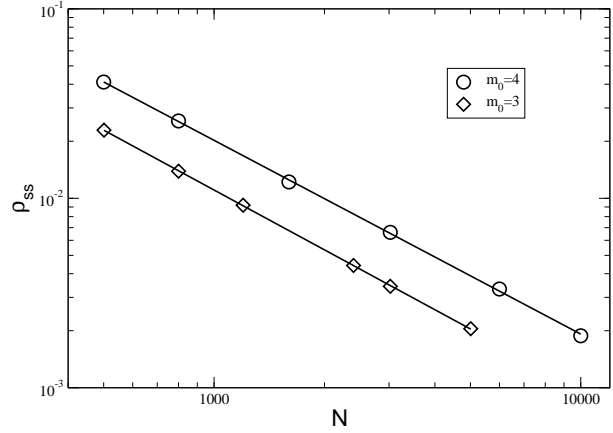}
\end{center}
\caption{
Plot of the behaviour of $\varrho_{ss}$ as a function of the system's size $N$ for
$m_0=3$ (diamonds) and $m_0=4$ (circles). The full lines are power-law fits $\sim N^{-\gamma}$,
with $\gamma=1.02\pm0.02$ ($m_0=4$) and $\gamma=1.05\pm0.01$ ($m_0=3$).}
\label{rhoss}
\end{figure}

The same behaviour is found for higher $m_0$, as one can see in Figure \ref{rhoss},
so that we can conclude stating the relation

\begin{equation}
\label{Rrhoss}
\varrho_{ss}(N;m_0)\sim N^{-1}\ \ \ \forall m_0
\end{equation}

Of course, last equation, together with (\ref{tstar1}), demonstrates  also that the time $t^*$ for leaving the Regime I
is proportional to $\log(N)$ for every value of $m_0$.

\subsection{Frozen state}
\label{RIII}

It is straightforward to understand that the steady state cannot last forever. Indeed,
according to Eq. (\ref{solmII}), the survived clusters averaged size should diverge after
a time $\bar t$ given by

\[
\bar t=\frac{N}{m_0\varrho_{ss}^2}\sim N^3
\]

\begin{figure}[h]
\begin{center}
\includegraphics[angle=0,width=8cm,clip]{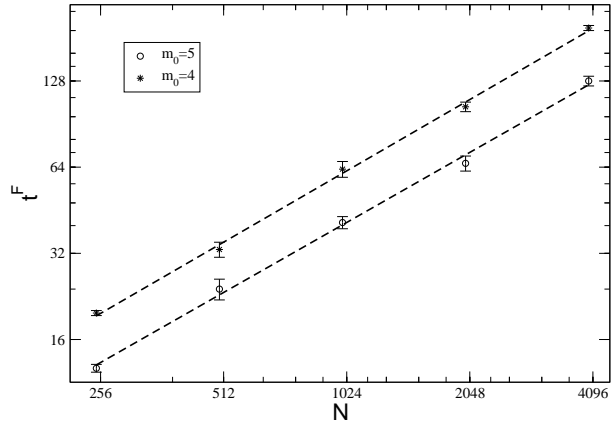}
\end{center}
\caption{
Plot of the behaviour of $t^F$ as a function of $N$ for $m_0=4$ (stars) and $m_0=5$
(circles); the dashed lines are power-law fits with exponent $\delta=0.83\pm0.02$ and
$\delta=0.81\pm0.02$, respectively.}
\label{tF}
\end{figure}

Anyway, it is obviously impossible that $m\rightarrow+\infty$, since of course
$m(t)\leq N$. In fact, the dynamics freezes well before this time $\bar t$: in
Figure \ref{tF} we report the behaviour of the freezing time for a couple of values
of $m_0$, from which it is possible to see that the freezing time $t^F$ follows
actually a power-law on $N$, but with exponent $\delta\approx0.8$ instead of $3$.

Here we wonder what kind of frozen state is finally reached by the system. Let
us consider the general equation (\ref{mgen}) ruling over the evolution of $m(t)$.
Assuming the sizes of survived clusters as independent from each other at
every time, so that we can write $\langle m_im_j\rangle=\langle m_i\rangle\langle
m_j\rangle=m^2\ \forall t$, we obtain

\begin{equation}
\label{mIII}
\frac{dm}{dt}=\frac{m^2(N-m)}{N^2}\varrho^2(t)
\end{equation}

By integrating last relation, we find now

\begin{equation}
\label{solmIII}
\frac{m(t)e^{-[N/m(t)]}}{N-m(t)}=K\cdot\exp{\left[\int_0^t\varrho^2(\tau)d\tau\right]}
\end{equation}

being $K$ a suitable (positive) constant. Now, in the limit $t\rightarrow+\infty$, there
is an instant $t^F$ (the freezing time we introduced above) such that $\varrho(t)=\varrho(t^F)$
$\forall t\geq t^F$, and said $\varrho^F$ this cooperators density of the frozen state,
from equation (\ref{solmIII}) it is clear that there are only two possible final configurations:

\begin{itemize}

\item (A) - If we have $\varrho^F=0$ (no cooperators in the final state), then it must
necessarily be $m^F<N$, that is the frozen state is disordered.

\

\item (B) - If instead we have $\varrho^F>0$ (finite fraction of cooperators in the final
state), then the integral at right side diverges, so that it must be $m^F=N$, thus
the frozen state is ordered ({\it i.e.\rm} only one survived cluster remains in the system).

\end{itemize}

The configuration (A) is completely lacking in cooperators, so, in order to
be frozen, the difference in size between two clusters whatever must be always less
than their distance minus $1$: in the opposite case, as one can see from the payoff
matrix (\ref{paymatr1}), there would be players who could become cooperators after
an interaction. On the other hand, the configuration (B) is pretty easy to
understand, since when the entire system is occupied by just one cluster, dynamics
stops by definition.
Now, in the steady state regime, the cooperators density is so small that it is
possible to get a fluctuation pushing the system in the disordered frozen state, with
no cooperators and many clusters in it. If instead such a fluctuation does not happen,
the normal dynamics given by equations (\ref{rhogen}) and (\ref{mgen}), or even more
simply by (\ref{solmIII}), will drive the system into the ordered frozen state,
with a finite density of cooperators,
and one mega-cluster occupying the whole system. For these reasons we expect that the
probability of the system to end in the disordered frozen configuration increases
with $\varrho_{ss}$ decreasing, {\it i.e.\rm} with $N$ increasing and $m_0$ decreasing. Actually,
for $m_0=1$ and after 1000 simulations, we observed the system ending in the ordered
state only three times for $N=100$, just once for $N=200$, and never for higher $N$.
On the other hand for $m_0\geq4$, we did not ever observe the system falling in the
disordered configuration, since in this case $\varrho_{ss}$ becomes small enough only at
very high $N$, when the freezing time is too big to be observed. Finally, the ratio
between the number of times in which the frozen state is ordered over the number in
which it is ordered drops from 0.86 for $N=200$ to 0.1 for $N=2000$
in the case $m_0=2$, and it is still 0.95 for $N=3024$ when $m_0=3$.

\begin{figure}[h]
\begin{center}
\includegraphics[angle=0,width=8cm,clip]{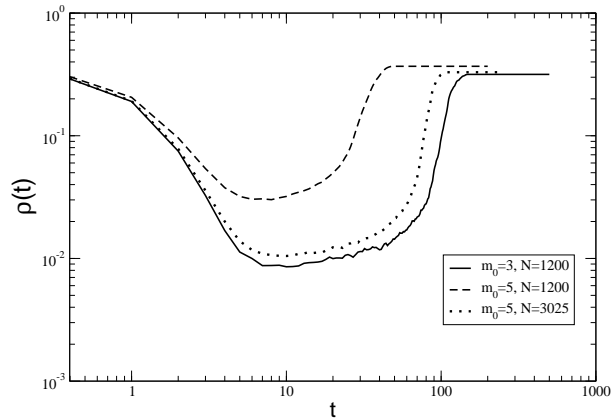}
\end{center}
\caption{
Plot of $\varrho(t)$ as a function of time for some values of $m_0$ and $N$, after
100 simulations. Every simulation ended in the completely ordered frozen state;
indeed the (not shown) shape of $m(t)$ for the values of parameters here reported
is the same of Figure \ref{gendyn}, {\it i.e.\rm} it is always $m^F/N=1$.}
\label{frznDens}
\end{figure}

An interesting aspect of the ordered configuration is that the density of cooperators
is in this case finite but less than one: so, even though the disordered frozen state
is just the Nash equilibrium $p_1^E=0$ of the SHM (see section \ref{hom}), the ordered one
is not the perfect counterpart of the analogous in SHM. That can be explained because
when the system is very close to the completely ordered state, the agents belonging to
the biggest cluster have no interest in cooperation, so that most of them will be defectors.
This is shown in Figures \ref{gendyn} and \ref{frznDens} where it is easy to see how the
abundance of cooperators in the frozen ordered state is always well smaller than $1/2$
(remaining around $1/3$). On the other hand, this is not a real Nash equilibrium, since
it does not exist in the thermodynamical limit.

\subsection{Limit of very large initial clusters}
\label{GVm}

Until now we have dealt with small values of the initial clusters size $m_0$: more
precisely, so far we have exploited the thermodynamical limit supposing fixed $m_0$ as
$N$ increases. Now, one could wonder what happens to the system if we set instead
$m_0=zN$ (with $0<z<1$) before doing the limit $N\rightarrow+\infty$. Indeed,
in the SHM, a transition between the phase with the unique stable Nash equilibrium
$p_1^E$ and the phase with two stable equilibria (in particular the new one
$p_2^E$) takes place for $m=x+1$, being $x$ the half averaged distance among all
clusters. An analogous transition in DHM somehow happens, but in a rather trivial
way: indeed, when $m_0$ diverges (even remaining much smaller than $N$), a single
interaction between two cooperators will create a new cluster very much bigger than
the others, thus the system will reach the ordered state soon, typically after much
less than 10 time units.

\section{Conclusions and Perspectives}
\label{concl}

In this paper we have depicted some new ideas for the study and the understanding
of the phenomena of social aggregation in human communities.
First, we suggested a theoretical treatment based on both statistical mechanics and
game theory.
Secondly, a fundamental feature of our approach is the interplay between the
inclination of every agent to cooperate with others (in order to live in
groups as big as possible), and an opposite attitude not to move away from
the actual group since joining a new one involves a work accomplished by the
agent itself. This work, needed by an individual when it associates to a stranger
cluster, is interpreted as a ``distance'' in the abstract space of cultural
coordinates: the more two groups have different CC, the more they are far away
from each other in this space, the more is the work an individual must spend to
go from one cluster to the other. In order to test the reliability of this approach,
we conceived a couple of very simple toy models, both constructed with the general features
described above, the first one being a pure evolutionary population model, the second one
an agent model with a well defined dynamics at a microscopical level.

The results obtained with such toy models suggest that cooperation is the most
suitable strategy only in presence of very big clusters, so that the gain in
fitness of the individuals who join these big groups is greater than the distance
they had to cover to reach their new ``accommodation''.
More precisely, using the language of game theory, we found that cooperation is an
evolutionary stable Nash equilibrium, when the averaged size of clusters is bigger
enough with respect to the averaged distance among them.

These models are of course a tough simplification of the real world, and contain some
unsatisfactory features: in particular, the property of the clusters to merge when
two cooperators of them meet is quite strong, and also the definition of distance
between clusters appears to be somehow arbitrary. Improving the models in these aspects
can be the goal of future researches. Anyway, despite such problems, our results
are qualitatively realistic for some important social phenomena which involve
human societies. Indeed, our results suggest that in an area occupied by a great
deal of small communities, distributed more or less uniformly, nobody has interest to
move from home to another community, since there is no real difference among the
communities, and a displacement would mean only a work to accomplish without any gain
in fitness. However, when some of these communities, because of a change in the external
conditions, or for a simple fluctuation, become quite big with respect to the other ones,
they assume the role of centres of attraction, destinations of the immigration of people
from anywhere, so that these centres reach soon the typical size of a metropolis. This
aggregation mechanism seems actually to be what really happened during several
urbanization phenomena through history, as for instance the ``urban explosion'' in the
basin of the Mediterranean Sea around the XII Century BC, or also in Western Europe during
the Industrial Revolution. It is worth to notice that in this picture the merging of
two groups when only two cooperators interact is not so unrealistic, since presumably
an immigrant will call and invite to the big city his former fellow citizens.
Moreover, in many cases this same dynamics is apparently at work when religions,
political parties, idioms or other kinds of social aggregations grow up inside a society.
Finally, it is worth also to mention the result that, as we saw in subsection \ref{RIII},
it is not really necessary that every individual has to cooperate in order to merge
different clusters into one: on the contrary, the fraction of cooperators can be
less than $0.5$ also in systems made up of only one big cultural cluster.

Of course, deeper studies and further interpretations are needed, but
the fact that so oversimplified models give already reasonable results is
very encouraging and suggest to continue with this kind of study.

\section*{Acknowledgements}

We thank F. Bagnoli and T. Carletti for the very useful suggestions they gave us during
the editing of this article.

\appendix
\section*{APPENDIX}

In order to determine the stability of the Nash equilibria shown by Eq. (\ref{equilibri})
it is enough to integrate the replicator equation given in (\ref{rplc}), obtaining

\begin{equation}
\label{APP_sol}
\frac{p^{1-\omega}(1-p)^{\omega}}{|p-\omega|}=G\cdot\exp\left[2x(\omega-1)t\right]
\end{equation}

being $G$ a positive integration constant. Now, solving explicitly in $p$ the relation
(\ref{APP_sol}) is in general impossible, but if we define the quantity
$L_{\omega}\in\overline{\mbox{\bf R\rm}}$ as

\begin{equation}
\label{APP_lom}
L_{\omega}\doteq\lim_{t\rightarrow+\infty}\exp\left[2x(\omega-1)t\right]
\end{equation}

it is easy to see that for $-2x<\varepsilon<0$, that is $\omega>1$, we have
$L_{\omega}=+\infty$ and this implies necessarily $p\rightarrow0^+$ (the exponent $1-\omega$
is naturally negative); on the other hand, for $\varepsilon<-2x$, that is $\omega<0$, it is
$L_{\omega}=0$, implying again necessarily $p\rightarrow0^+$ (now the exponent $1-\omega$
is positive). So, it is proven that only $p_1^E$ is stable for $\varepsilon<0$.
Finally, for $\varepsilon>0$, that is $0<\omega<1$, it results $L_{\omega}=0$, with both
exponents $\omega$ and $1-\omega$ positive, implying that $p$ can tend to $0^+$ or $1^-$,
but not to $p_3^E=\omega$, which is therefore unstable.

\

Regarding the evaluation of the size of the basins of attraction in the case
$\varepsilon>0$, we could effectively compute the smallest fluctuation needed to escape
from each of them. Anyway, it is for sure easier, watching Figure \ref{phasediagr},
if we consider that the unstable equilibrium $p_3^E$ is the separator between the two basins,
so that the basin of stability of $p_1^E$ is of course of size $\omega=2x/(2x+\varepsilon)$,
while the basin of $p_2^E$ has size $1-\omega=\varepsilon/(2x+\varepsilon)$. Besides,
such basins will be equal for $\varepsilon^*=2x$, {\it i.e.\rm} for $m^*=3x+1$.
\end{document}